\documentclass[a4paper, conference]{IEEEtran}

\usepackage{cite}
\usepackage{amsmath,amssymb,amsfonts}
\usepackage{algorithmic}
\usepackage{graphicx}
\usepackage{textcomp}
\usepackage{xcolor}
\usepackage{paralist}
\usepackage{eurosym}
\usepackage{flushend}
\usepackage{url}
\usepackage[utf8]{inputenc}
\usepackage[colorlinks=true]{hyperref}
\usepackage{soul}
\usepackage{caption}
\usepackage[list=true]{subcaption}
\usepackage{booktabs}
\usepackage{tabu}
\usepackage{wrapfig}
\usepackage{arydshln}
\usepackage[normalem]{ulem}
\usepackage{multirow}
\usepackage{siunitx}
\usepackage{adjustbox}
\usepackage{array}
\usepackage{diagbox}
\usepackage{tikz}
\usepackage{tabularx}
\usepackage{xfrac}
\usetikzlibrary{shapes.geometric, arrows.meta, positioning}

\newcommand\blfootnote[1]{%
  \begingroup
  \renewcommand\thefootnote{}\footnote{#1}%
  \addtocounter{footnote}{-1}%
  \endgroup
}

\begin{document}

\title{The S-ICDF Dataset: Sionna-Simulated Dynamic Interference Characterization and Direction Finding}

\author{\IEEEauthorblockN{Christian Wielenberg, Lucas Heublein, Jonathan Ott, Alexander Mattick, Nisha L. Raichur, Jonas Pirkl, \\ Lukas Schelenz, Tobias Feigl, George Yammine, Christopher Mutschler, Felix Ott}
  \IEEEauthorblockA{Fraunhofer Institute for Integrated Circuits IIS, 90411 Nürnberg, Germany}
  \IEEEauthorblockA{\{christian.wielenberg, tobias.feigl, george.yammine, christopher.mutschler, felix.ott\}@iis.fraunhofer.de}
}

\maketitle

\begin{abstract}
Jamming and spoofing threaten wireless and satellite navigation by disrupting or manipulating radio frequency (RF) signals, undermining availability, integrity, and trust. Robust interference monitoring (i.e., detection, classification, characterization, and direction finding) is therefore essential to identify and localize anomalous signals. While machine learning (ML) promises improved performance in complex environments, its development and validation depend on large-scale datasets that capture realistic signal and channel variability. Collecting such data in the real world is difficult because intentional jamming is illegal and ground-truth attribution is confounded by propagation, hardware, and environmental effects. To address this gap, we create and publish S-ICDF, a large-scale indoor interference dataset generated with Sionna, a GPU-accelerated simulation library for physical-layer wireless communications. S-ICDF covers $102$ interference configurations, including diverse antenna array patterns, bandwidths, and simulation settings such as noise level and reflection depth. We further provide baseline results by benchmarking S-ICDF with classical estimation and direction finding (DF) methods (MUSIC, ESPRIT, and CAPON) and with modern ML approaches. The dataset is publicly available at: \href{https://gitlab.cc-asp.fraunhofer.de/darcy_gnss/sicdf_dataset}{https://gitlab.cc-asp.fraunhofer.de/darcy\_gnss/sicdf\_dataset}
\end{abstract}
\IEEEpeerreviewmaketitle

\section{Introduction}
\label{label_introduction}

Jamming devices interfere with global navigation satellite system (GNSS) signals emitted by RF sources and represent a serious threat, undermining the robustness and reliability of both precise positioning and wireless communications~\cite{ferre_fuente,swinney_woods,pirayesh_zeng}. This problem has grown severe in recent years as readily obtainable jammers have become increasingly common~\cite{mehr_minetto_dovis,miguel_chen_lo}. As a result, interference detection, classification~\cite{ott_heublein_icl}, characterization~\cite{heublein_feigl_crpa}, localization~\cite{raichur_ion_gnss,heublein_wielenberg}, and mitigation~\cite{chen_liu_huang} have emerged as key research areas. Numerous solutions have been explored, spanning traditional signal processing techniques~\cite{yang_kang,murrian_narula,gross_humphreys} as well as ML-based methods~\cite{raichur_heublein,heublein_feigl_jispin}.

However, training such ML-based methods (particularly with the emergence of large foundation models) demands large-scale datasets that not only match the target task, but also comprehensively cover and remain well-balanced across all parameters the model is intended to learn, ensuring that relevant conditions and variations are adequately represented rather than dominated by a small subset of frequent cases~\cite{he_garcia}. In the context of interference monitoring, this entails broad and uniform coverage of jammer characteristics, including waveform/type (e.g., continuous wave, noise, swept, pulsed), center frequency and bandwidth, temporal behavior (duty cycle, burst/sweep dynamics), transmit power or interference-to-noise-ratio (INR), and spatial/propagation conditions (e.g., location and mobility, (non-)line-of-sight (LoS/NLoS), and multipath~\cite{heublein_feigl_crpa}. Collecting real-world recordings involving jamming is often legally restricted~\cite{jammer_enforcement}, and typically necessitates extensive measurement campaigns, such as the Jammertest in Norway~\cite{broumandan_pirsiavash}.

Hence, our objective is to generate a simulated dataset and to benchmark ML models on this dataset to enable rapid model iteration and development. Although publicly available (simulated) datasets exist~\cite{humphreys_bhatti,swinney_woods_data,rahman_bhuiyan,ferre_richter,richter_ferre_lohan,gomez_casco_crosta,albright_powers}, they typically exhibit at least one of the following limitations: (1) insufficient coverage of a broad range of interference characteristics; (2) the absence of multi-patch antenna measurements required by DF methods; or (3) stationary receiver setups that preclude time-series-based DF approaches. To overcome these shortcomings, we employ Sionna~\cite{hoydis_cammerer} as the underlying simulation environment.

\setlength{\intextsep}{6pt}
\setlength{\columnsep}{12pt}
\begin{wrapfigure}{R}{4.4cm}
    \begin{minipage}[b]{1.0\linewidth}
        \centering
        \vspace{-0.1cm}
        \includegraphics[trim=76 38 36 50, clip, width=1.0\linewidth]{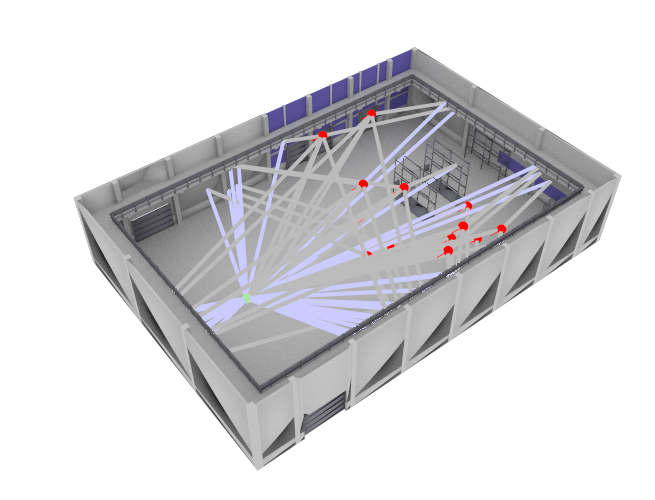}
        \caption{Sionna simulation with ray tracing.}
        \label{figure_sionna_simulation}
    \end{minipage}
\end{wrapfigure}

\textbf{Contributions.} In the following, we summarize our main contributions: (1) In Sionna, we model an indoor industrial environment and employ ray tracing to simulate propagation from an interference source and the resulting signals received by an antenna array (refer to Figure~\ref{figure_sionna_simulation}). (2) We consider $110$ distinct interference parameterizations, targeting both interference characterization and direction-of-arrival (DoA) estimation between the source and the receiver. The receiver is modeled as a $2{\times}2$-multi-patch array providing raw in-phase and quadrature (IQ) samples. (3) To support time-series-based DF via a synthetic aperture effect, the antenna platform moves through the environment during data acquisition. (4) We evaluate interference characterization and DF performance using both classical signal processing baselines and ML-based approaches. The S-ICDF dataset is released as a public benchmark to facilitate reproducible evaluation and comparison in future research, offering broad and controlled coverage of interference conditions, array-based IQ measurements, and realistic time-varying multipath due to ray-traced propagation and receiver motion.

\begin{table*}[t]
    \centering
    \caption{Overview of publicly available real-world and simulated GNSS interference datasets.}
    \label{table_gnss_interference_datasets}
    \setlength{\tabcolsep}{2pt}
    \begin{tabular}{p{3.2cm} p{3.4cm} p{2.2cm} p{3.9cm} p{4.7cm}}
        \toprule
        \textbf{Dataset} & \textbf{Real-world vs.~Simulated} & \textbf{Interference Type} & \textbf{Data Modality} & \textbf{Direction Finding Suitability} \\
        \midrule
        TEXBAT~\cite{humphreys_bhatti} & Real-world (field recordings) & Spoofing & IQ recordings & Primarily single-channel \\
        Oak Ridge (OAKBAT)~\cite{albright_powers} & Real-world (digitized) & Spoofing & Digitized RF/IQ recordings & Limited (mainly single-channel) \\
        L.I.N.K.~Hall~\cite{heublein_wielenberg} & Real-world (indoor) & Moving jammer & IQ recordings & $2{\times}2$-array for DF, not phase-coherent \\
        Tuni2025 (TG-GGSD)~\cite{rahman_bhuiyan} & Real-world (lab) & Spoofing & IQ recordings + scenario metadata & Single stream; not DF-centric \\
        In-lab DF Validation~\cite{ferre_richter} & Real-world (lab) & Jamming & Measurement-based evaluation & Moderate (DF evaluated) \\
        GATEMAN~\cite{richter_ferre_lohan} & Real-world (in-lab) & Jamming & Wideband IQ recordings & Limited \\
        Evil WaveForms~\cite{gomez_casco_crosta} & Simulated (GNSS simulator) & Intf.~waveforms & Simulated IQ waveforms & Limited (no array/motion labels for DF) \\
        Raw IQ Dataset~\cite{swinney_woods_data} & Simulated & Jamming & Labeled IQ samples & No array/multi-sensor information \\
        \bottomrule
    \end{tabular}
\end{table*}

\section{Related Work}
\label{label_related_work}

Table~\ref{table_gnss_interference_datasets} provides an overview of publicly available datasets for GNSS interference monitoring differing substantially in how threats are generated and what they enable: TEXBAT~\cite{humphreys_bhatti}, OAKBAT~\cite{albright_powers}, and Tuni2025~\cite{rahman_bhuiyan} provide scenario-based RF/IQ recordings primarily targeting spoofing analysis and benchmarking, whereas the IQ jamming classification dataset~\cite{swinney_woods_data} offers labeled samples geared toward supervised jammer-type recognition. For jamming-focused studies, the GATEMAN release~\cite{richter_ferre_lohan} supplies wideband in-lab recordings of GNSS and jammer signals, complemented by the experimental validation study in~\cite{ferre_richter} that evaluates detection and DF-methods under controlled conditions. The Evil WaveForms work~\cite{gomez_casco_crosta} is simulator-based and provides synthetic threat waveforms for reproducible testing, but lacks the array and motion information required for robust DF/localization benchmarking. Heublein et al.~\cite{heublein_wielenberg} presented a large-scale dataset of moving GNSS jamming devices recorded in an indoor industrial environment with dynamic multipath, providing multi-patch raw IQ measurements together with ground-truth relative position labels for jammer DF and localization.

While useful as benchmarks, most datasets either lack the synchronized multi-antenna phase information or the trajectory/pose ground truth needed to train and evaluate DF under moving antennas or jammers. Moreover, motion diversity and key dynamic effects (time-varying multipath, LoS/NLoS changes, and broad jammer parameter/INR coverage) are often limited, restricting robust generalization and detailed comparisons. To address these limitations and to enable controlled, scalable generation of motion-rich, fully labeled array measurements under realistic propagation, we simulate our dataset using Sionna~\cite{hoydis_cammerer}. Prior work has primarily used Sionna for large-scale synthetic channel and radio-map generation in digital-twin settings, whereas we are, to the best of our knowledge, the first to leverage Sionna to generate time-series raw IQ measurements for DF with a moving antenna array in the presence of interference sources.

Research on GNSS interference monitoring is advanced, with a wide range of well-established signal processing and ML-based techniques. Accordingly, we utilize well-established baselines to ensure comparable evaluation: for characterization tasks, we employ an XceptionTime~\cite{rahimian_zabihi} architecture, while for DF we consider classical estimators, including MUSIC~\cite{schmidt}, ESPRIT~\cite{roy_kailath}, and CAPON~\cite{hassanien_shahbazpanahi}, which are commonly used reference methods for DoA estimation.
\section{Simulation}
\label{label_simulation}

In this section, we present the complete simulation-to-benchmark workflow. We first describe the end-to-end data-generation pipeline (Sec.~\ref{label_simulation_pipeline}). We then detail the Sionna-based simulation setup, including the ray-tracing environment and all relevant configuration parameters (Sec.~\ref{label_simulation_sionna}). Finally, we introduce the S-ICDF dataset and summarize its structure, parameter ranges, and data splits (Sec.~\ref{label_simulation_dataset}).

\subsection{Pipeline}
\label{label_simulation_pipeline}

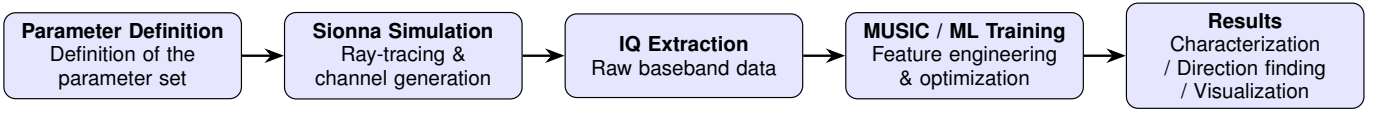
\begin{figure*}
    \centering
    \resizebox{\textwidth}{!}{%
    \begin{tikzpicture}[
        node distance=0.5cm and 0.5cm,
        block/.style={
            rectangle, 
            draw, 
            fill=blue!10, 
            text width=2.6cm, 
            align=center, 
            rounded corners, 
            minimum height=1.0cm,
            font=\sffamily\scriptsize
        },
        arrow/.style={
            thick, 
            ->, 
            >=Stealth
        }
    ]
    \node (param) [block] {\textbf{Parameter Definition} \\ \scriptsize Definition of the parameter set};
    \node (sionna) [block, right=of param] {\textbf{Sionna Simulation} \\ \scriptsize Ray-tracing \& channel generation};
    \node (iq) [block, right=of sionna] {\textbf{IQ Extraction} \\ \scriptsize Raw baseband data};
    \node (model) [block, right=of iq] {\textbf{MUSIC / ML Training} \\ \scriptsize Feature engineering \& optimization};
    \node (results) [block, right=of model] {\textbf{Results} \\ \scriptsize Characterization / Direction finding / Visualization};
    \draw [arrow] (param) -- (sionna);
    \draw [arrow] (sionna) -- (iq);
    \draw [arrow] (iq) -- (model);
    \draw [arrow] (model) -- (results);
    \end{tikzpicture}
    }
    \vspace{-0.5cm}
    \caption{Pipeline from parameter definition and Sionna simulation to IQ extraction, MUSIC/ML training, and evaluation.}
    \label{fig:Pipeline}
    \vspace{-0.1cm}
\end{figure*}

The simulation pipeline is designed to systematically quantify how individual simulation parameters affect both the received signal structure and the performance of direction finding algorithms. Figure~\ref{fig:Pipeline} provides an overview. First, a parameter configuration is specified by defining the range and resolution of each variable of interest. To enable an interpretable sensitivity analysis, we vary only a single parameter at a time while holding all remaining parameters fixed at their default values, such that any performance variation can be attributed to that parameter. Each configuration is then provided to Sionna, which models the transmission environment (including propagation and channel effects) and generates corresponding complex baseband signals (see Section~\ref{label_simulation_sionna}). From the simulator output, we extract raw IQ samples, preserving the full complex-valued information required by downstream localization methods (see Section~\ref{label_simulation_dataset}). The resulting IQ sequences are subsequently processed by the algorithms under study, including classical baseline algorithms and the considered ML models trained directly on the simulated data (see Section~\ref{label_experiments}). Finally, performance metrics are computed and aggregated across parameter settings, enabling a systematic comparison of the impact of each parameter on MUSIC, ESPRIT, CAPON and ML-based localization performance.

\subsection{Sionna Simulation Environment}
\label{label_simulation_sionna}

NVIDIA Sionna~\cite{hoydis_cammerer} is an open-source, GPU-accelerated (TensorFlow-based) simulation library for wireless PHY/link-level research where transmitters, channels, and receivers are built as modular, differentiable blocks so the user can run fast Monte-Carlo evaluations and also backpropagate gradients end-to-end for learning-based designs. Sionna includes detailed components for modern chains (e.g., OFDM/MIMO, coding, estimation/detection) and an optional ray-tracing module (Sionna RT) to generate geometry/material-aware channels from 3D scenes.

The simulation environment represents an industrial hall-like scenario and is designed to capture the dense multipath conditions typical of indoor industrial settings; an example realization is shown in Figure~\ref{figure_sionna_simulation}. The scene is implemented using Sionna’s ray-tracing engine, which computes physically grounded propagation paths between transmitter and receiver. It comprises a large rectangular layout with predominantly metallic boundary surfaces (walls, ceiling, and floor) and includes interior structures such as shelving units that serve as reflectors and scatterers. This geometry yields a large number of multipath components, providing a challenging yet realistic testbed for localization methods. The maximum number of reflections considered per propagation path is governed by the ray-tracing reflection depth, enabling controlled variation of multipath richness across simulation runs.

\subsection{The S-ICDF Dataset}
\label{label_simulation_dataset}

This section presents the S-ICDF dataset, which is designed to facilitate a detailed assessment of how individual signal and channel parameters influence IQ signal structure and the performance of signal localization algorithms.

\paragraph{Transmitter \& Receiver Setup} A single signal source is placed at a fixed location within the scene and transmits at a center frequency of $1.57542\,\text{GHz}$, corresponding to the GPS-L1 band center frequency. The transmitted waveform is determined by a configurable source-signal parameter and can be selected from multiple signal types. The receiver is modeled as a uniform antenna array with either a $2 \times 2$ planar geometry or a $8 \times 1$ linear geometry. During acquisition, the array traverses the hall while maintaining a fixed orientation relative to the transmitter. Inter-element spacing is specified by the antenna-distance parameter, and the element gain characteristics are defined by the selected antenna gain pattern. The received signal at each antenna element is sampled over a simulation bandwidth of $100\,\text{MHz}$, resulting in $1{,}024$ complex IQ samples per snapshot.

\setlength{\intextsep}{6pt}
\setlength{\columnsep}{12pt}
\begin{wrapfigure}{R}{3.2cm}
    \begin{minipage}[b]{1.0\linewidth}
        \centering
        \vspace{-0.2cm}
        \includegraphics[trim=26 3 3 30, clip, width=1.0\linewidth]{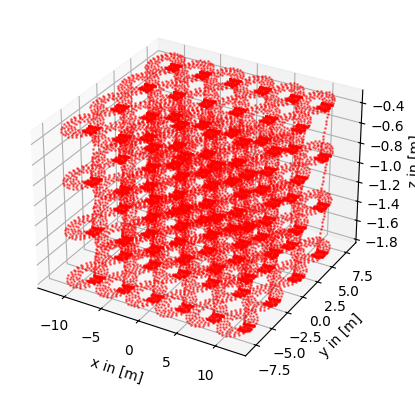}
        \caption{Trajectory of the antenna.}
        \label{fig_trajectory}
    \end{minipage}
\end{wrapfigure}

\paragraph{Parameter Simulation Methodology} The dataset is generated according to a one-at-a-time (OAT) parameter-variation strategy. In each experiment, only a single parameter of interest is varied over its predefined range, while all remaining parameters are fixed at their respective default values. This design enables a clear and unambiguous attribution of performance changes to the parameter under investigation, without confounding effects arising from simultaneous multi-parameter variation. An overview of all simulation, interference, and antenna parameters is provided in Table~\ref{table_sionna_parameters}. For each parameter variation, we generate a complete positioning dataset comprising 216,089 samples, which is subsequently divided into an 80/20 train--test split. Each sample corresponds to a specific point along a receiver trajectory, as illustrated in Figure~\ref{fig_trajectory}.

\begin{table}[t!]
    \centering
    \caption{Overview of all simulation, interference, and antenna receiver parameters.}
    \label{table_sionna_parameters}
    \vspace{-0.1cm}
    \begin{tabular}{l r r}
        \toprule
        \multicolumn{3}{c}{\textbf{Simulation Environment Parameters}} \\
        \multicolumn{2}{c}{\textbf{Parameter}} & \multicolumn{1}{c}{\textbf{Value}} \\
        \midrule
        \multicolumn{2}{l}{Bandwidth of simulated IQ samples} & \multicolumn{1}{r}{$100\,\text{MHz}$} \\
        \multicolumn{2}{l}{Center frequency of simulated IQ samples} & \multicolumn{1}{r}{$1.57542\,\text{GHz}$} \\
        \multicolumn{2}{l}{Received signal length} & \multicolumn{1}{r}{$1{,}024$} \\
        \multicolumn{2}{l}{Number of simultaneous interference sources} & \multicolumn{1}{r}{$1$} \\ 
        \multicolumn{2}{l}{Trajectory length of each recording (number of samples)} & \multicolumn{1}{r}{$216{,}089$} \\
        \toprule
        \multicolumn{3}{c}{\textbf{Interference \& Antenna Setting Parameters}} \\
        \multicolumn{1}{c}{\textbf{Parameter}} & \multicolumn{1}{c}{\textbf{Range}} & \multicolumn{1}{c}{\textbf{Default}} \\
        \midrule
        Source bandwidth & $1\,\text{MHz}$ to $20\,\text{MHz}$ & $20\,\text{MHz}$ \\
        SNR & $-20\,\text{dB}$ to $20\,\text{dB}$ & $20\,\text{dB}$ \\
        Source signal & Chirp, Modulated, Noise,  & Chirp \\
        & Multitone, FrequHopper & \\
        Antenna distance & $[0.05\,\text{m}, 0.09\,\text{m}, 0.095\,\text{m}, 0.2\,\text{m}]$ & $0.09\,\text{m}$ \\
        Reflection depth & $[2, 5, 7]$ & $5$ \\
        Refraction & [\texttt{True}, \texttt{False}] & \texttt{True} \\
        Array layout & [$2 \times 2$ or $8 \times 1$] & $2 \times 2$ \\
        Antenna gain pattern & [dipole, iso, hw-dipole, tr38901] & dipole \\
        \bottomrule
    \end{tabular}
    \vspace{-0.3cm}
\end{table}

\begin{figure}[!t]
    \centering
    \begin{minipage}[t]{0.324\linewidth}
        \centering
        \includegraphics[trim=10 10 10 10, clip, width=1.0\linewidth]{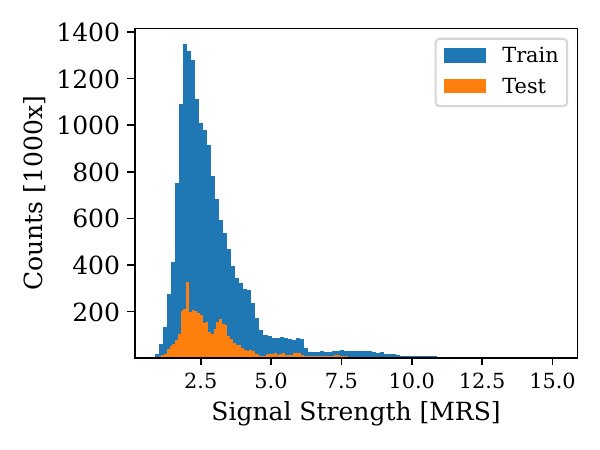}
        \vspace{-0.4cm}
        \subcaption{Signal strength.}
        \label{figure_data_overview_signal_strength}
    \end{minipage}
    \hfill
    \begin{minipage}[t]{0.324\linewidth}
        \centering
        \includegraphics[trim=10 10 10 10, clip, width=1.0\linewidth]{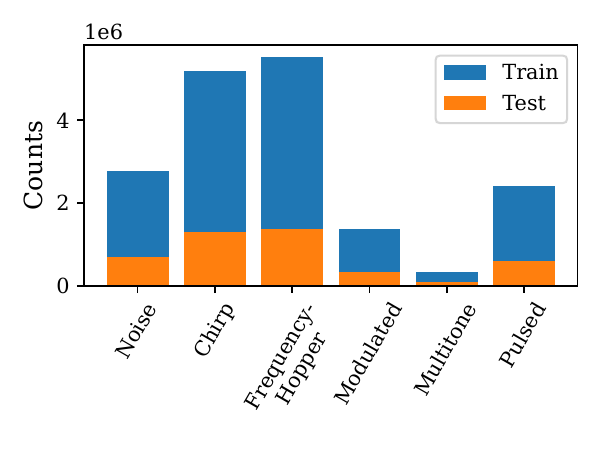}
        \vspace{-0.4cm}
        \subcaption{Interference types.}
        \label{figure_data_overview_samples}
    \end{minipage}
    \hfill
    \begin{minipage}[t]{0.324\linewidth}
        \centering
        \includegraphics[trim=10 10 10 10, clip, width=1.0\linewidth]{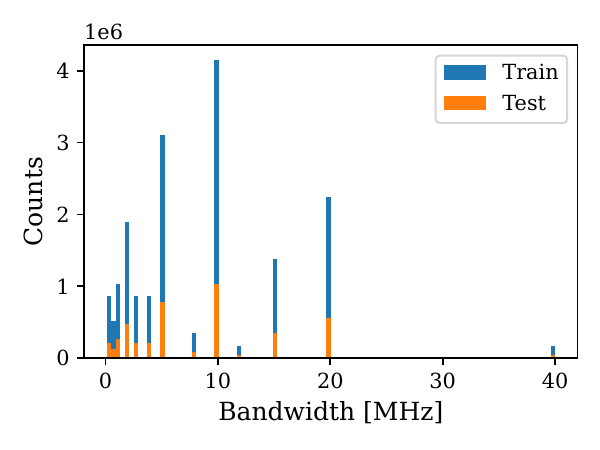}
        \vspace{-0.4cm}
        \subcaption{Bandwidths.}
        \label{figure_data_overview_bandwidth}
    \end{minipage}
    \vspace{-0.15cm}
    \caption{Histograms of signal sharacteristics.}
    \label{figure_data_overview}
    \vspace{-0.3cm}
\end{figure}

\paragraph{Parameter Description} The following parameters are varied individually over their respective ranges, while all remaining parameters are fixed at their default values, as summarized in Table~\ref{table_sionna_parameters}. The \textit{antenna element spacing} is evaluated at four discrete values, namely $0.05\,\text{m}$, $0.09\,\text{m}$, $0.095\,\text{m}$, and $0.2\,\text{m}$, with a default value of $0.09\,\text{m}$. At a center frequency of $1.57542\,\text{GHz}$, this corresponds to inter-element spacings ranging from $0.26\lambda$ to $1.1\lambda$. We chose both $0.09\,\text{m}$ and $0.095\,\text{m}$ as we have observed ambiguities for a wavelength of exactly $0.5\lambda$. For the \textit{array layout}, two configurations are considered: a $2 \times 2$ rectangular array and an $8 \times 1$ linear array. These configurations provide two fundamentally different aperture geometries, thereby affecting angular resolution and spatial diversity. Four antenna \textit{gain patterns} are considered. The isotropic pattern serves as an idealized baseline with uniform sensitivity in all directions. The dipole and half-wave dipole patterns introduce directional sensitivity characteristic of practical antenna elements. In addition, the 3GPP TR~38.901 model is included as a standardized antenna pattern. The \textit{signal-to-noise} (SNR) parameter controls the level of additive noise applied to the received signal and is varied from $-20\,\text{dB}$ to $20\,\text{dB}$ in steps of $2\,\text{dB}$. Low SNR values correspond to severely degraded reception conditions, whereas high SNR values represent near-ideal conditions. Figure~\ref{figure_data_overview_signal_strength} shows the histogram of receiver signal power (mean root squared over IQ samples). The \textit{reflection depth} determines the maximum number of reflections that a propagation path may undergo within the Sionna ray-tracing engine and is evaluated for values of $2$, $5$, and $7$. Larger values result in a denser and more complex multipath environment, thereby increasing simulation realism at the expense of computational cost. \textit{Refraction} is modeled as a Boolean parameter that enables or disables refractive propagation effects. When set to \texttt{True}, signals are permitted to propagate through penetrable surfaces, yielding additional propagation paths; when set to \texttt{False}, only reflective propagation paths are considered. This parameter isolates the impact of refractive multipath components on localization performance.

\begin{figure}[!t]
\captionsetup[subfigure]{font=footnotesize}
    \centering
    \begin{minipage}[t]{0.155\linewidth}
        \centering
        \includegraphics[trim=46 14 84 10, clip, width=1.0\linewidth]{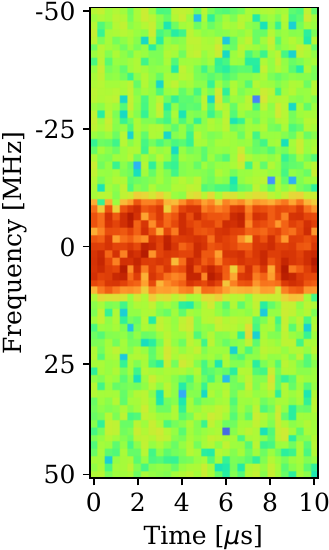}
        \vspace{-0.4cm}
        \subcaption{Noise.}
        \label{figure_spectrograms_iq1}
    \end{minipage}
    \hfill
    \begin{minipage}[t]{0.155\linewidth}
        \centering
        \includegraphics[trim=46 14 84 10, clip, width=1.0\linewidth]{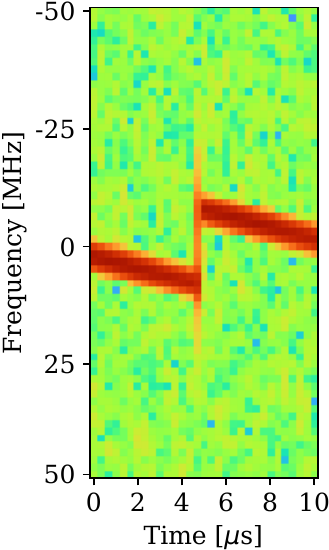}
        \vspace{-0.4cm}
        \subcaption{Chirp.}
        \label{figure_spectrograms_iq2}
    \end{minipage}
    \hfill
    \begin{minipage}[t]{0.155\linewidth}
        \centering
        \includegraphics[trim=46 14 84 10, clip, width=1.0\linewidth]{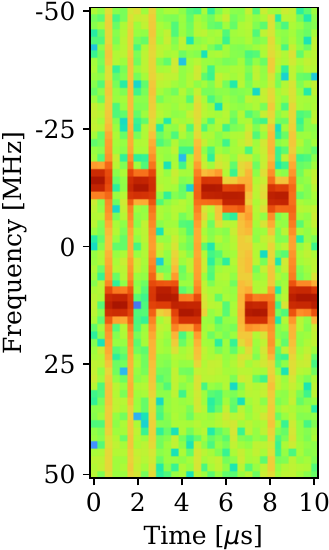}
        \vspace{-0.4cm}
        \subcaption{Hopper.}
        \label{figure_spectrograms_iq3}
    \end{minipage}
    \hfill
    \begin{minipage}[t]{0.155\linewidth}
        \centering
        \includegraphics[trim=46 14 84 10, clip, width=1.0\linewidth]{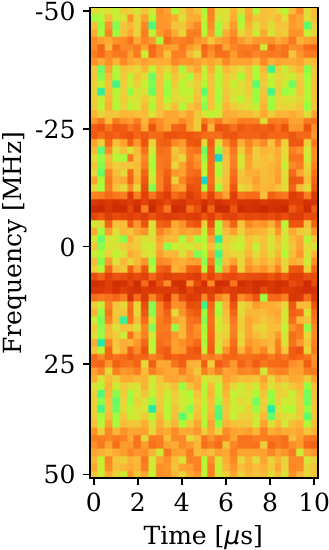}
        \vspace{-0.4cm}
        \subcaption{Modul.}
        \label{figure_spectrograms_iq4}
    \end{minipage}
    \hfill
    \begin{minipage}[t]{0.155\linewidth}
        \centering
        \includegraphics[trim=46 14 84 10, clip, width=1.0\linewidth]{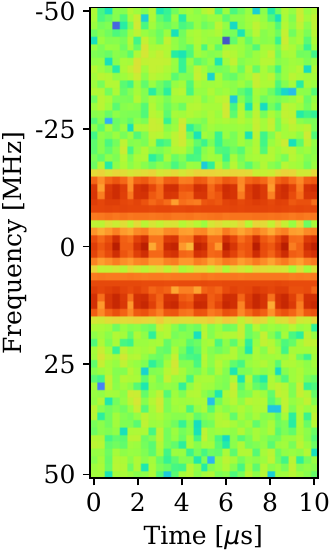}
        \vspace{-0.4cm}
        \subcaption{Multit.}
        \label{figure_spectrograms_iq5}
    \end{minipage}
    \hfill
    \begin{minipage}[t]{0.155\linewidth}
        \centering
        \includegraphics[trim=46 14 84 10, clip, width=1.0\linewidth]{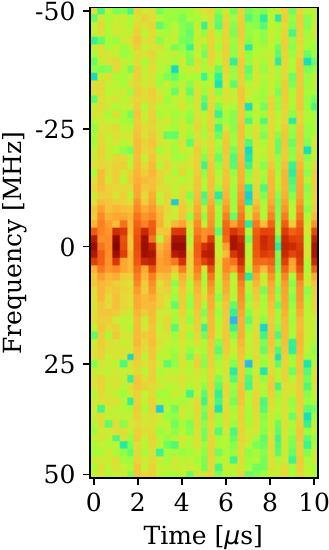}
        \vspace{-0.4cm}
        \subcaption{Pulsed.}
        \label{figure_spectrograms_iq6}
    \end{minipage}
    \vspace{-0.15cm}
    \caption{Spectrograms for six main interference modulations.}
    \label{figure_spectrograms_iq}
    \vspace{-0.2cm}
\end{figure}

\paragraph{Signal Modulation} The \textit{source signal} defines the transmitted waveform; here, we use predefined generated vector signals rather than simple synthetic waveforms. In total, \(102\) signal files are included, covering six distinct signal classes and thus a broad range of interference waveform types. Figure~\ref{figure_spectrograms_iq} shows examples of the six main interference modulations, while Figure~\ref{figure_data_overview_samples} presents their distribution in the dataset. For \textit{Chirp} signals, we consider linear (Lin) and parabolic (Para) sweep profiles with fast (Fa) (\(0.025\,\text{ms}\)), medium (Med) (\(0.25\,\text{ms}\)), and slow (Sl) (\(2.5\,\text{ms}\)) sweep times, and vary the bandwidth from \(2\) to \(20\,\text{MHz}\). \textit{FrequencyHopper} rapidly switch across discrete carrier frequencies, with bandwidth and dwell time as the main varying parameters. \textit{Noise} signals are spectrally flat emissions with varying occupied bandwidth; additionally, binary offset carrier (BOC)-modulated variants are included to impose characteristic spectral structure. For \textit{Multitone}, multiple equally spaced sinusoidal tones are transmitted simultaneously, with the number of tones and total occupied bandwidth determining the spectral density. For \textit{Modulated}, modulation scheme, symbol rate, and occupied bandwidth are varied to capture diverse structured spectral/temporal characteristics. Finally, \textit{Pulsed} signals alternate between transmission and silence with a fixed \(1{:}1\) signal-to-silence ratio, while pulse repetition interval and bandwidth are varied. A histogram of the interference bandwidths is given in Figure~\ref{figure_data_overview_bandwidth}.
\section{Experiments}
\label{label_experiments}

\paragraph{DF with Classical Methods} The primary objective of this dataset is to assess the impact of signal parameters on DF performance. To this end, we establish a baseline using well-known classical methods, namely MUSIC~\cite{schmidt}, ESPRIT~\cite{roy_kailath}, and CAPON~\cite{hassanien_shahbazpanahi}. The DF algorithms are evaluated on the $20\%$ test portion of the simulated data, and the mean azimuth and elevation errors are used as benchmark metrics. For MUSIC and ESPRIT, the evaluation grid spans \(0^\circ\) to \(360^\circ\) in azimuth and \(0^\circ\) to \(90^\circ\) in elevation, with an angular resolution of \(1^\circ\) in both dimensions.

\paragraph{Characterization \& DF with ML} Our objective is to characterize all $102$ interference modulation types. For each of the $4 \times 6$ circle-grid trajectories, we simulate distinct interference parameterizations. We partition the dataset into an 80/20 train--test split and train an XceptionTime~\cite{rahimian_zabihi} model on raw IQ inputs of size $1{,}024 \times 4 \times 2$, where $1{,}024$ is the time dimension, $4$ is the array dimension, and $2$ is the real and imaginary part of the IQ-signal. The dataset comprises $440{,}821$ training samples and $110{,}172$ test samples. The model is trained to predict the interference characteristics, namely the main interference class, modulation type, and bandwidth, using a cross-entropy loss, and to regress the azimuth and elevation angles as well as the source--receiver distance using a mean squared error (MSE) loss. The network outputs a $128$-dimensional feature representation that is fed into a linear prediction head. We use a batch size of $64$, train for $100$ epochs, and use the standard SGD optimizer with an initial learning rate of $10^{-4}$, applying a multi-step schedule that reduces the learning rate by a factor of $0.1$ after $60$ and $80$ epochs. The overall objective is
\begin{equation}
    \mathcal{L}_{\text{total}} = \lambda_1 \mathcal{L}_\text{class} + \lambda_2 \mathcal{L}_\text{char} + \lambda_3 \,\mathcal{L}_\text{az} + \lambda_4 \,\mathcal{L}_\text{el} + \lambda_5 \,\mathcal{L}_\text{dis},
\end{equation}
where we set $\lambda_1 = \lambda_2 = 1$, and $\lambda_3 = \lambda_4 = \lambda_5 = 0.3$.

\begin{figure}[!t]
    \centering
    \begin{minipage}[t]{0.492\linewidth}
        \centering
        \includegraphics[trim=10 24 16 50, clip, width=1.0\linewidth]{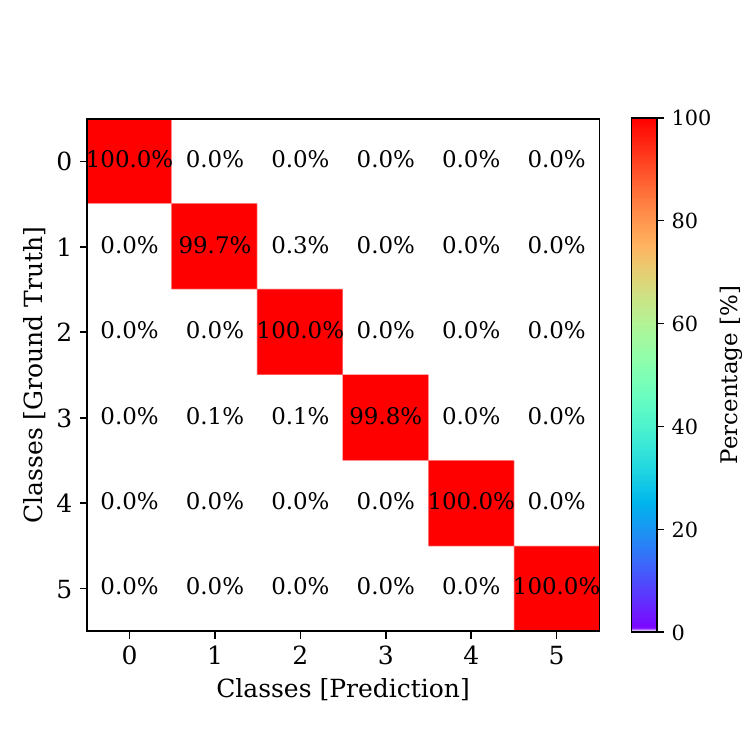}
    \end{minipage}
    \hfill
    \begin{minipage}[t]{0.492\linewidth}
        \centering
        \includegraphics[trim=9 68 64 100, clip, width=1.0\linewidth]{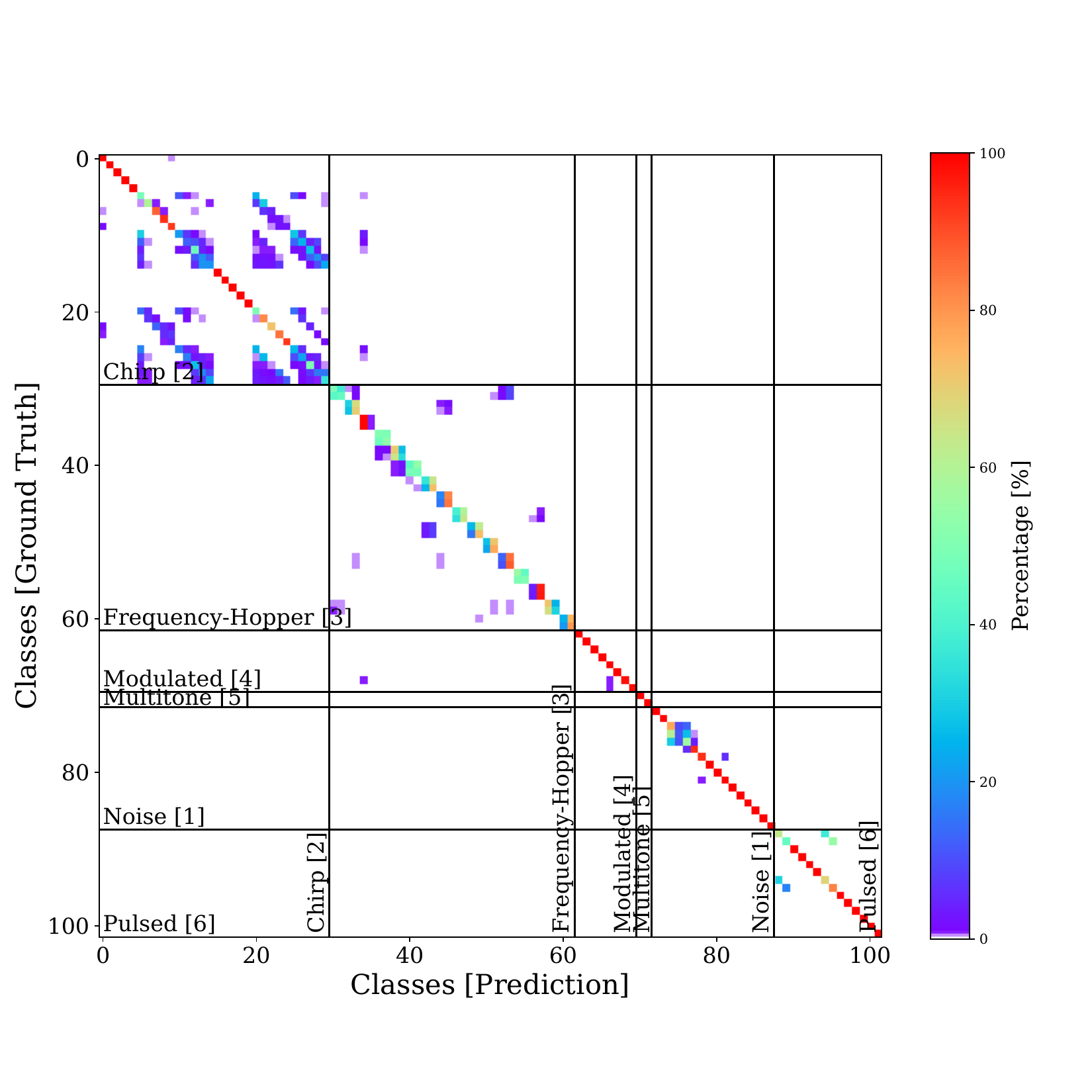}
    \end{minipage}
    \caption{Evaluation of classification (left) and characterization (right) tasks with XceptionTime~\cite{rahimian_zabihi} on raw IQ samples.}
    \label{figure_results_class_charact}
\end{figure}

\section{Evaluation}
\label{label_evaluation}

Consistently, we report the MSE of relative position (in m), azimuth (in $^\circ$), and elevation (in $^\circ$), and the accuracy (in \%) of classification and characterization.

\begin{figure*}[!t]
    \centering
    \begin{minipage}[t]{0.245\linewidth}
        \centering
        \includegraphics[trim=6 6 6 6, clip, width=1.0\linewidth]{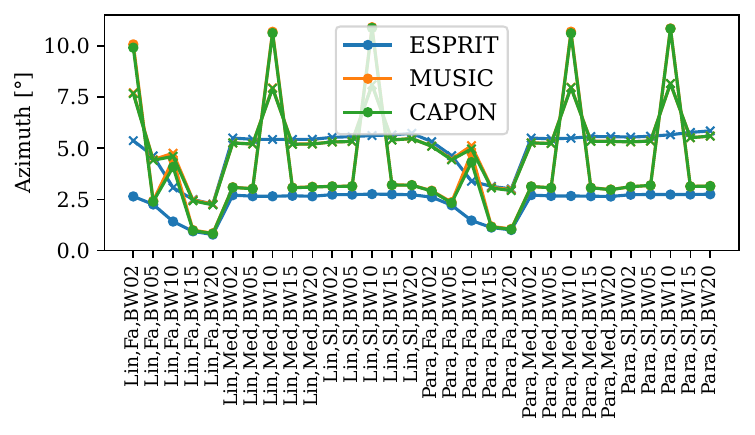}
        \vspace{-0.5cm}
        \subcaption{Chirp.}
        \label{figure_df_modulation1}
    \end{minipage}
    \hfill
    \begin{minipage}[t]{0.245\linewidth}
        \centering
        \includegraphics[trim=6 6 6 6, clip, width=1.0\linewidth]{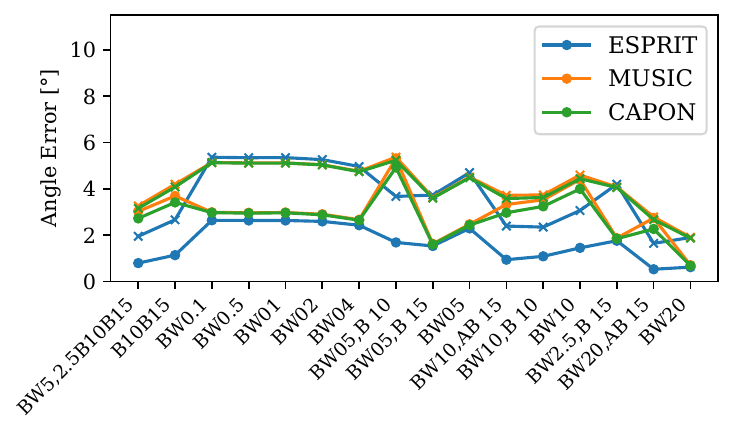}
        \vspace{-0.5cm}
        \subcaption{Noise.}
        \label{figure_df_modulation2}
    \end{minipage}
    \hfill
    \begin{minipage}[t]{0.245\linewidth}
        \centering
        \includegraphics[trim=6 6 6 6, clip, width=1.0\linewidth]{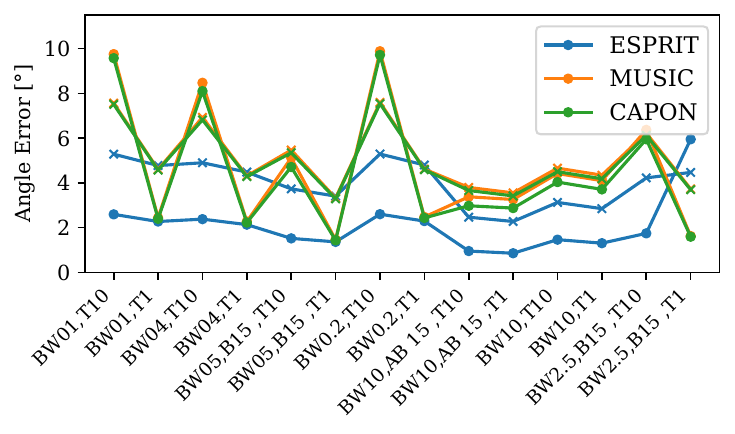}
        \vspace{-0.5cm}
        \subcaption{Pulsed.}
        \label{figure_df_modulation3}
    \end{minipage}
    \hfill
    \begin{minipage}[t]{0.245\linewidth}
        \centering
        \includegraphics[trim=6 6 6 6, clip, width=1.0\linewidth]{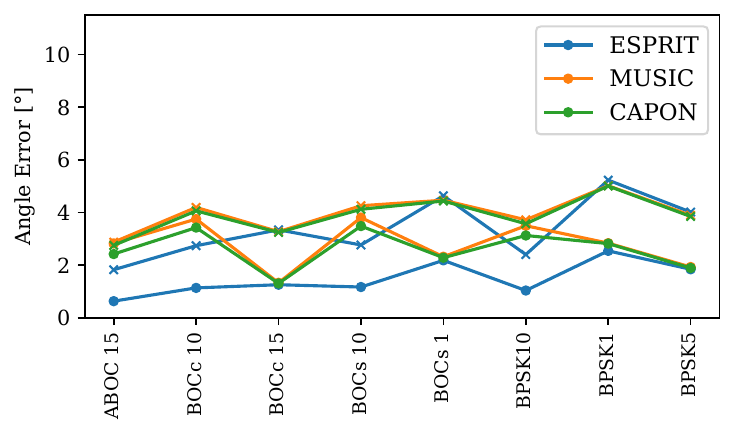}
        \vspace{-0.5cm}
        \subcaption{Modulated.}
        \label{figure_df_modulation4}
    \end{minipage}
    \vspace{-0.2cm}
    \caption{Results of the azimuth (dots) and elevation (cross) orientation errors (in $^{\circ}$, MSE) for different interference modulations. ``BW'' denotes the bandwidth, ``B'' denotes the width of the BOC separation, ``T'' denotes the pulse rate in \text{ms}.}
    \label{figure_df_modulation}
    \vspace{-0.3cm}
\end{figure*}

\paragraph{Modulation Characterization} Figure~\ref{figure_results_class_charact} shows the results for interference classification and characterization. At the class level, which leads to an accuracy of 99.89\%, \textit{Noise}, \textit{Multitone}, and \textit{Pulsed} are identified with high accuracy, as reflected by the strong diagonal structure. Most remaining errors occur between \textit{Chirp} and \textit{FrequencyHopper}, whose signatures become similar at small bandwidths. The \textit{Modulated} class is also occasionally confused with these two categories, although only to a limited extent. For the characterization task, which leads to an accuracy of 71.99\%, the majority of errors occur within the same interference types rather than across different classes. In particular, \textit{Chirp} signals are mainly confused between linear and parabolic sweeps and between medium and slow sweep rates, while \textit{FrequencyHopper} signals are primarily misclassified across similar bandwidth settings. Within the \textit{Noise} category, \textit{AltBOC} and \textit{BOC} variants exhibit similar spectral patterns, which leads to additional ambiguity.

\begin{figure}[!t]
    \centering
    \includegraphics[trim=6 6 6 6, clip, width=0.72\linewidth]{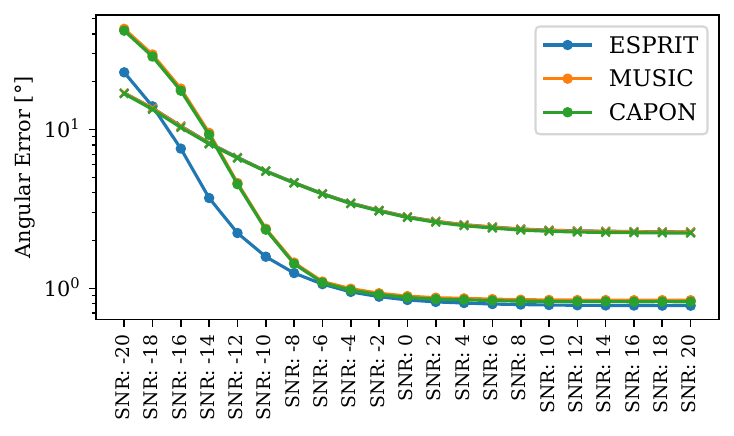}
    \caption{Evaluation results of the azimuth orientation error (in $^{\circ}$, MSE, logarithmic) for various SNR [in dB].}
    \label{label_figure_snr_results}
    \vspace{-0.3cm}
\end{figure}

\paragraph{Classical DF} Figure~\ref{figure_df_modulation} compares the DF performance of MUSIC, ESPRIT, and CAPON across interference modulations. Overall, ESPRIT achieves the most accurate and stable azimuth estimates, consistently outperforming MUSIC and CAPON across most waveform variants. The clearest advantage is observed for \textit{Chirp} and \textit{Modulated} signals, where ESPRIT maintains comparatively low azimuth errors while MUSIC and CAPON exhibit larger fluctuations. For \textit{Noise} signals, the performance gap between the methods is smaller, indicating that these waveforms are generally easier to localize. In contrast, \textit{Pulsed} signals appear to be the most challenging, as all methods show increased variability and higher errors for several parameterizations. A further observation is that azimuth errors are generally lower than elevation errors, suggesting that azimuth estimation is less strongly affected by waveform-dependent ambiguities and multipath. In summary, the figure indicates that the interference modulation has a noticeable impact on DF accuracy, with ESPRIT providing the most robust overall performance.

\paragraph{Evaluation of SNR} Figure~\ref{label_figure_snr_results} shows the azimuth DF error of MUSIC, ESPRIT, and CAPON as a function of the SNR over the range $[-20\,\text{dB}, 20\,\text{dB}]$. The SNR considered here varies within a single environment, as path-loss effects are explicitly accounted for in the simulation. A clear threshold behavior can be observed: below $-8\,\text{dB}$, the azimuth error increases strongly for all methods, whereas above $-8\,\text{dB}$ the error decreases rapidly and saturates at a low value. For low-SNR, additive noise increasingly masks the phase and amplitude relations between antenna elements, which degrades the covariance-matrix estimate and makes subspace separation less reliable. As the SNR increases, the signal component becomes more dominant, leading to more stable direction estimates and lower azimuth errors. At high SNR, the remaining error is mainly limited by factors such as multipath propagation, array geometry, and finite sample effects rather than by noise.

\begin{table}[t]
    \centering
    \caption{Results for different Sionna parameter settings. We evaluate azimuth ($\epsilon_a$) and elevation ($\epsilon_e$) errors (in $^{\circ}$, MSE).}
    \label{table_results_overview}
    \vspace{-0.1cm}
    \setlength{\tabcolsep}{2.5pt}
    \begin{tabular}{ p{0.5cm} p{0.5cm} p{0.5cm} p{0.5cm} p{0.5cm} p{0.5cm} p{0.5cm} }
        \toprule
        & \multicolumn{2}{c}{\textbf{MUSIC}} & \multicolumn{2}{c}{\textbf{ESPRIT}} & \multicolumn{2}{c}{\textbf{CAPON}} \\
        \multicolumn{1}{c}{\textbf{Parameter}} & \multicolumn{1}{c}{$\epsilon_a$} & \multicolumn{1}{c}{$\epsilon_e$} & \multicolumn{1}{c}{$\epsilon_a$} & \multicolumn{1}{c}{$\epsilon_e$} & \multicolumn{1}{c}{$\epsilon_a$} & \multicolumn{1}{c}{$\epsilon_e$} \\
        \midrule
        \multicolumn{7}{l}{\textbf{Default parameter setting:} Array layout: $2{\times}2$, Gain: dipole,} \\
        \multicolumn{7}{l}{Reflection depth = 5, Refraction = \texttt{False}, Spacing: $0.09\,\text{m}$} \\
        \multicolumn{1}{l}{} & \multicolumn{1}{r}{$0.84$} & \multicolumn{1}{r}{$2.26$} & \multicolumn{1}{r}{$0.78$} & \multicolumn{1}{r}{$2.27$} & \multicolumn{1}{r}{$0.82$} & \multicolumn{1}{r}{$2.23$} \\
        \midrule
        \multicolumn{1}{l}{Array layout: $8{\times}1$} & \multicolumn{1}{r}{$13.19$} & \multicolumn{1}{r}{ - } & \multicolumn{1}{r}{$13.19$} & \multicolumn{1}{r}{ - } & \multicolumn{1}{r}{$2.60$} & \multicolumn{1}{r}{ - } \\
        \multicolumn{1}{l}{Gain: iso} & \multicolumn{1}{r}{$0.83$} & \multicolumn{1}{r}{$2.26$} & \multicolumn{1}{r}{$0.76$} & \multicolumn{1}{r}{$2.26$} & \multicolumn{1}{r}{$0.81$} & \multicolumn{1}{r}{$2.22$} \\
        \multicolumn{1}{l}{Gain: hw\_dipole} & \multicolumn{1}{r}{$0.85$} & \multicolumn{1}{r}{$2.27$} & \multicolumn{1}{r}{$0.79$} & \multicolumn{1}{r}{$2.27$} & \multicolumn{1}{r}{$0.83$} & \multicolumn{1}{r}{$2.23$} \\
        \multicolumn{1}{l}{Gain: TR 38.901} & \multicolumn{1}{r}{$1.86$} & \multicolumn{1}{r}{$3.51$} & \multicolumn{1}{r}{$1.46$} & \multicolumn{1}{r}{$3.51$} & \multicolumn{1}{r}{$1.62$} & \multicolumn{1}{r}{$3.46$} \\
        \multicolumn{1}{l}{Reflection depth = 2} & \multicolumn{1}{r}{$2.35$} & \multicolumn{1}{r}{$3.53$} & \multicolumn{1}{r}{$1.70$} & \multicolumn{1}{r}{$3.61$} & \multicolumn{1}{r}{$2.08$} & \multicolumn{1}{r}{$3.43$} \\
        \multicolumn{1}{l}{Refraction = \texttt{True}} & \multicolumn{1}{r}{$0.75$} & \multicolumn{1}{r}{$1.98$} & \multicolumn{1}{r}{$0.67$} & \multicolumn{1}{r}{$1.98$} & \multicolumn{1}{r}{$0.74$} & \multicolumn{1}{r}{$1.96$} \\
        \multicolumn{1}{l}{Antenna distance: $0.05\,\text{m}$} & \multicolumn{1}{r}{$1.22$} & \multicolumn{1}{r}{$47.95$} & \multicolumn{1}{r}{$1.16$} & \multicolumn{1}{r}{$5.41$} & \multicolumn{1}{r}{$1.20$} & \multicolumn{1}{r}{$5.33$} \\
        \multicolumn{1}{l}{Antenna distance: $0.095\,\text{m}$} & \multicolumn{1}{r}{$1.66$} & \multicolumn{1}{r}{$7,45$} & \multicolumn{1}{r}{$0.78$} & \multicolumn{1}{r}{$2.27$} & \multicolumn{1}{r}{$1.57$} & \multicolumn{1}{r}{$2.23$} \\
        \multicolumn{1}{l}{Antenna distance: $0.2\,\text{m}$} & \multicolumn{1}{r}{$102.31$} & \multicolumn{1}{r}{$32.56$} & \multicolumn{1}{r}{$77.13$} & \multicolumn{1}{r}{$61.45$} & \multicolumn{1}{r}{$71.48$} & \multicolumn{1}{r}{$29.96$} \\
        \bottomrule
    \end{tabular}
    \vspace{-0.3cm}
\end{table}

\paragraph{Evaluation of Sionna Parameters} Table~\ref{table_results_overview} shows that the dataset captures meaningful dependencies on key parameters. Under default settings, all three methods perform well, with azimuth errors around $0.8^{\circ}$ and elevation errors around $2.2^{\circ}$. ESPRIT consistently outperforms MUSIC and CAPON. The antenna \textit{gain pattern} has a moderate effect: dipole, isotropic, and half-wave dipole yield nearly identical results, whereas the 3GPP TR 38.901 pattern noticeably degrades both azimuth and elevation accuracy. A higher \textit{reflection depth} leads to a more robust DF, while enabling \textit{refraction} slightly improves all methods, likely because propagation paths provide more stable spatial information. The default \textit{antenna spacing} of $0.09\,\text{m}$ gives the best overall performance. At $0.095\,\text{m}$ MUSIC and CAPON have spatial ambiguities between the estimated directions. Similarly, for the $8 \times1 $ array configuration with a normalized element spacing of $0.5\lambda$, spatial ambiguities are observed, where the estimated source direction is occasionally mirrored at broadside. To account for this symmetric ambiguity, azimuth estimates for the $8 \times 1$ array are normalized to $\left[-90^{\circ},\, 90^{\circ}\right]$.

\begin{figure}[!t]
\captionsetup[subfigure]{font=footnotesize}
    \centering
    \begin{minipage}[t]{0.324\linewidth}
        \centering
        \includegraphics[trim=10 10 10 10, clip, width=1.0\linewidth]{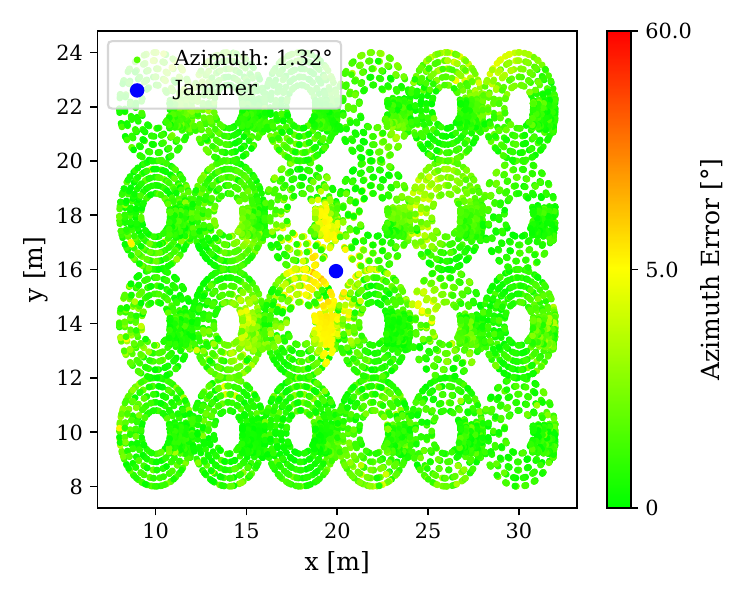}
        \vspace{-0.5cm}
        \subcaption{Chirp.}
        \label{figure_results_resnet_azimuth1}
    \end{minipage}
    \hfill
    \begin{minipage}[t]{0.324\linewidth}
        \centering
        \includegraphics[trim=10 10 10 10, clip, width=1.0\linewidth]{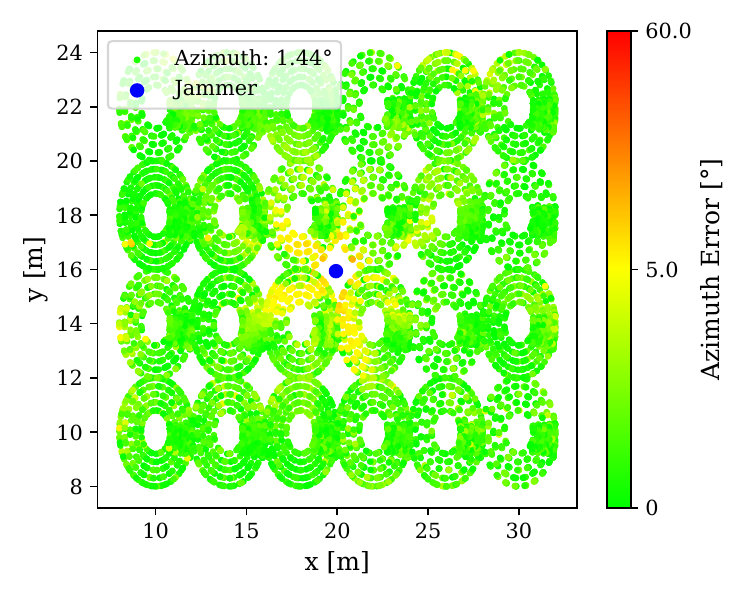}
        \vspace{-0.5cm}
        \subcaption{FrequHopper.}
        \label{figure_results_resnet_azimuth2}
    \end{minipage}
    \hfill
    \begin{minipage}[t]{0.324\linewidth}
        \centering
        \includegraphics[trim=10 10 10 10, clip, width=1.0\linewidth]{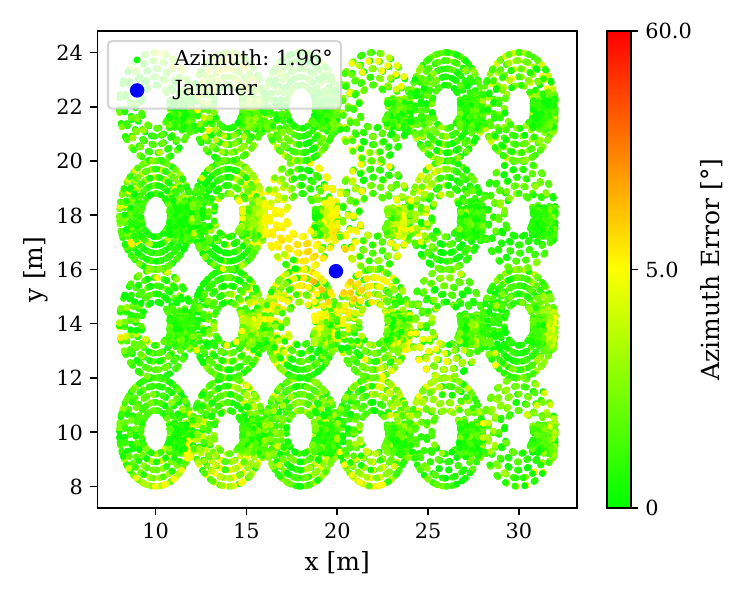}
        \vspace{-0.5cm}
        \subcaption{Noise/Pulsed.}
        \label{figure_results_resnet_azimuth3}
    \end{minipage}
    \vspace{-0.15cm}
    \caption{Evaluation of the azimuth prediction error.}
    \label{figure_results_resnet_azimuth}
    \vspace{-0.3cm}
\end{figure}

\paragraph{ML-Based DF} The final DF performance of the XceptionTime model achieves a mean error of $1.85^\circ$ in azimuth, $2.33^\circ$ in elevation, and $0.26\,\text{m}$ in position. Figure~\ref{figure_results_resnet_azimuth} illustrates the azimuth prediction error across all trajectory points for three representative interference types. Overall, the azimuth error tends to increase when the receiver is located closer to the interference source. Among the shown examples, \textit{Noise} and \textit{Pulsed} interference yields the highest azimuth error at $1.96^\circ$, whereas \textit{Chirp} interference results in the lowest error at $1.32^\circ$, indicating greater robustness of the model for this waveform class. This trend is consistent with the observations made for the classical DF methods.
\section{Conclusion}
\label{label_conclusion}

We introduced S-ICDF, a large-scale Sionna-based dataset for interference characterization and DF with controlled parameter variation. The benchmark demonstrates that S-ICDF captures meaningful dependencies on waveform, SNR, multipath, and antenna design, enabling systematic sensitivity analysis of both classical and ML-based methods. In our evaluation, XceptionTime achieves 99.89\% interference classification accuracy and competitive DF performance, while ESPRIT provides the most robust classical baseline ($0.78^{\circ}$).
\blfootnote{\textbf{Acknowledgments.} This work has been carried out within the PaiL project, funding code 50NP2506, sponsored by the German Federal Ministry for Transport (BMV) and supported by the German Space Agency at DLR, the Bundesnetzagentur (BNetzA), and the Federal Agency for Cartography and Geodesy (BKG).}

\bibliography{IPIN2026}
\bibliographystyle{IEEEtran}

\end{document}